# An Ingestible Light Source for Deep Photoacoustic Imaging

David C. Garrett and Lihong V. Wang

*Caltech Optical Imaging Laboratory, Andrew and Peggy Cherng Department of Medical Engineering, Department of Electrical Engineering, California Institute of Technology, Pasadena, CA 91125, USA*

\* Corresponding author: lvw@caltech.edu

**Abstract**

Photoacoustic tomography leverages ultrasound's deep tissue penetration to retrieve optical absorption contrast well beyond the optical diffusion limit. Conventional photoacoustic systems rely on externally delivered light and are therefore constrained by optical attenuation, limiting imaging depths to several centimeters. Here, we overcome this constraint using a compact, acoustically powered device that provides optical excitation directly from within the target medium. By exploiting the weak attenuation of low-MHz ultrasound, acoustic energy is transmitted through tissue to wirelessly power a pulsed laser diode. The emitted light pulses generate photoacoustic signals that encode local optical absorption at clinically relevant depths, which could enable imaging in regions such as the gastrointestinal tract that are inaccessible to surface-based illumination. We demonstrate this approach by imaging through a 12 cm thick phantom, establishing a pathway toward deep-tissue photoacoustic imaging.

**Introduction**

Since the early 2000s, capsule endoscopy (CE) has provided a minimally invasive means of imaging the gastrointestinal (GI) tract [1], [2], offering direct views of small intestine segments that upper or lower endoscopy cannot reach [3]. Compared with conventional endoscopy, CE is less invasive, does not require sedation, and enables prolonged monitoring as the capsule transits the GI tract [4], [5]. Ingestible CE devices contain a battery-powered camera and light source, and the images are transmitted wirelessly using radiofrequency signals to receivers positioned outside the body. However, since CE relies on conventional white-light photography, it reveals only the mucosal tissue surface [6]. Pathologies that reside within or beyond the bowel wall, such as transmural inflammation, angiodysplasia, or submucosal tumors, may not be detected [7]. An imaging modality capable of probing through the GI tract wall is therefore needed to improve the diagnosis and management of conditions like Crohn's disease, vascular malformations, and abdominal cancers [8].

MRI and CT provide excellent structural images of the abdomen, and endoscopic [9] and capsule-based [10] ultrasound systems can image beyond the mucosal wall. However, none of these modalities offer the functional and molecular contrast provided by optical imaging, such as visualization of blood vessel density. Photoacoustic tomography (PAT) is advantageous in imaging optical contrast at cm-scale depths beyond the tissue surface. Using high pulse-energy laser sources (~1 J), imaging depths of several centimeters have been obtained in regions like the breast [11], [12], [13]. These images reveal rich optical contrast relating to, for instance, blood oxygenation. However, to image regions deeper in the body like the GI tract using external illumination, PAT is limited by the effective attenuation coefficient $\mu_{\text{eff}} \sim 0.9 - 1.74 \text{ cm}^{-1}$ [11], [14]. At depths of several centimeters, the optical fluence is highly attenuated and results in prohibitively weak photoacoustic signals.

To perform PAT deeper in the body, photoacoustic endoscopy has been developed, where optical excitation is guided into the GI tract using optical fibers [15], [16]. The resulting acoustic signals are detected on the same endoscopic tether and are used for image reconstruction. However, tethered photoacoustic endoscopes are similarly constrained by the anatomical reach of conventional upper or lower endoscopes. Furthermore, these systems would require patient intubation and sedation, limiting their suitability for routine screening or longitudinal monitoring. There is therefore a gap in minimally-invasive technologies for imaging beyond the mucosal surface through the entire GI tract.

Here, we introduce an approach that enables wireless PAT (WPAT) at acoustically scalable depths well beyond the optical diffusion length. This device housing has a form factor comparable to existing CE devices (26 mm length, 11 mm diameter). PAT typically employs high-energy tabletop lasers, but here we perform PAT using a compact laser diode. Since MHz-scale acoustic waves propagate much deeper into the body, we use an external ultrasound transmitter to power the device and record the generated photoacoustic signals with a human-scale receiver array (60 cm in diameter). To synchronize optical excitation with ultrasound detection, electromagnetic pulses are received by the device and are used to trigger a pulsed 905 nm laser diode. We demonstrate WPAT by imaging a target through a 12 cm thick agar phantom.

# Methods

The WPAT device consists of three modules (Figure 1): optical excitation, acoustic power rectification, and electromagnetic triggering. Optical excitation is performed using a 905 nm laser diode, where ~0.6 µs, ~130 µJ pulses are discharged using a capacitor bank and a high-current field-effect transistor (FET). We choose this wavelength based on its deep tissue penetration and laser diode availability for LiDAR applications. Wireless power transfer is done acoustically using a 500 kHz receiving transducer, followed by a Cockcroft-Walton voltage multiplier circuit to maintain uniform input impedance while charging the capacitor bank. We trigger the laser firing using electromagnetic pulses, where a conformal helical antenna receives signals which are then rectified and fed to a monostable multivibrator circuit to generate consistent pulse durations. The resulting photoacoustic signals propagate through ~30 cm of water and are detected by a 60 cm diameter, 512-element ultrasound receiver array, which we have used for *in vivo* whole cross-sectional human imaging [17].

Each module is fabricated on a separate 0.4 mm thick printed circuit board (PCB), and they are interconnected using board-to-board wiring. The PCB design files are provided in the Supplementary Information. We model the device dimensions from commercial CE devices like the Medtronic Pillcam, CapsoVision CapsoCam, and Olympus EndoCapsule [18], [19]. This small size severely constrains the available components and requires several design choices to maximize performance despite the small device volume. Furthermore, the power budget for the design is strongly constrained by acoustic power transfer. We seal the enclosure using polydimethylsiloxane (PDMS) due to its transparency at 905 nm [20], similar acoustic impedance to water and tissue [21], and durability and biocompatibility in ingestible applications [22].

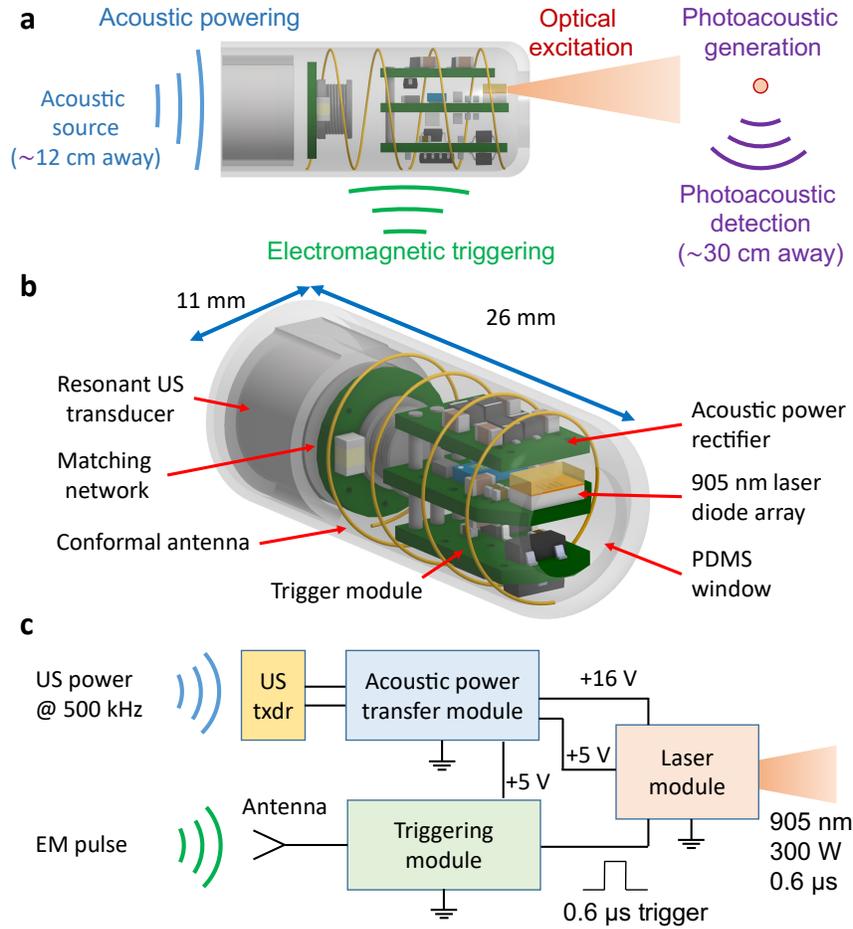

Figure 1. Wireless photoacoustic tomography device. **a** Overview of the energy sources used for powering, triggering, and exciting photoacoustic signals deep in tissue. **b** Model of the assembled device. **c** Block diagram of the device modules.

*Optical excitation module*

The laser excitation module (Fig. 2) consists of a pulsed 905 nm surface-mount laser diode array (Excelitas TPGAD1S11A-4A) driven for high peak-power operation. All four emitters are fired simultaneously to achieve ~300 W peak optical output during ~0.6 µs pulses. The diode is switched using a low-inductance gallium nitride FET (EPC2015), controlled by a gate driver (LMG1020), while a capacitor bank pre-charged to 15 V supplies the required current (~30 A peak). To achieve sharp optical pulses, we position the laser diode between two rows of capacitors totaling 4.5 µF where lines of vias are used to reduce the inductance during rapid high-current discharging. These capacitors are charged during the acoustic power transfer phase and are partially discharged during excitation.

We recorded the average optical pulse energy as 130 µJ using a photodiode power sensor (Thorlabs S120C). We compare this with the electrical energy consumed, which is found by the voltage drop on the charged capacitor bank from 15 V to ~9 V, where the electrical pulse energy is $E_{LD} = \frac{1}{2}C(V_i^2 - V_f^2) \sim 324$ µJ. This results in an electrical-to-optical efficiency of ~40%.

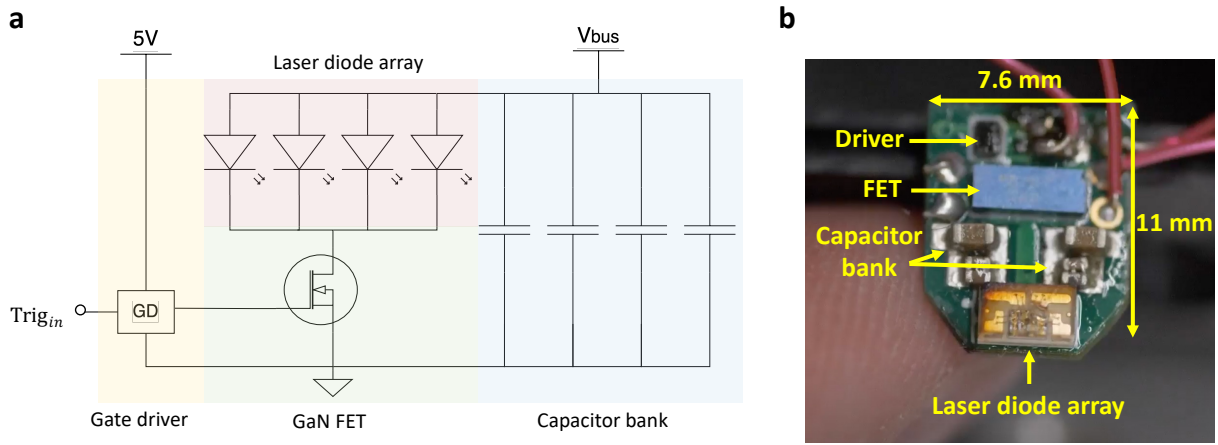

Figure 2. Optical excitation module. **a** Simplified circuit diagram. GD: gate driver. GaN: gallium nitride. **b** Assembled module.

*Acoustic power transfer*

With an average per-pulse electrical energy consumption of ~324 µJ, operating the laser at a 50 Hz repetition rate requires ~16 mW average electrical power. The quiescent power consumption of the other device components is < 1 mW. An internal battery could enable short-term operation, but the capsule must remain functional throughout its multi-hour GI transit, so relying on battery power alone would limit imaging duration. The required power is greater than what typical inductive or RF-based wireless power transfer systems report for deep tissue implants [23]. We instead take advantage of the low attenuation and guided energy delivery enabled by MHz-scale ultrasound. Ultrasound-based wireless power transfer has been demonstrated to yield mW-scale power to deeply implanted mm-sized devices [24], [25]. Furthermore, a single external acoustic array may in the future perform both ultrasound power delivery and photoacoustic detection, simplifying the clinical workflow while enabling shared calibration methods like sound speed correction that enhance both powering and imaging performance.

To wirelessly supply the ~16 mW average power required for laser operation, we first consider the acoustic safety limits. The FDA specifies a spatial peak temporal average intensity $I_{SPTA}$ < 720 mW/cm$^2$, which we treat as the maximum acoustic intensity that can occur in tissue or on the capsule surface, and a mechanical index (MI) < 1.9 MPa/$\sqrt{MHZ}$. Given our acoustic receiver surface area of 0.28 cm$^2$, up to ~200 mW of acoustic power can therefore be incident on the receiver. Achieving the required electrical output therefore requires an acoustic-to-electric conversion efficiency > 8%. Note that we optimize here for the total received electrical power when constrained by the safety standards rather than the efficiency of the entire acoustic link.

The chosen operating frequency is a trade-off between energy penetration and device size. Lower frequencies reduce tissue attenuation but require thicker transducers (for thickness-mode resonance) and, for a given peak rarefactional pressure, yield a higher mechanical index. We select 500 kHz as a balanced compromise. To efficiently convert acoustic energy into electrical power, we developed a custom 500 kHz compact resonant receiver. PZT-5A was selected as the piezoelectric material for its high electromechanical coupling factor ($k_{33}$ = 0.72). The total acoustic-to-electrical power efficiency at resonance is governed by acoustic and electrical impedance matching and is ultimately limited by internal mechanical loss and dielectric loss, with higher $k_{33}$ generally supporting greater achievable efficiency [26]. The parameter $k_{33}$ characterizes the strength of electromechanical coupling and is related to the separation between resonance and antiresonance, rather than directly specifying acoustic-to-electrical power efficiency. In practice, the delivered electrical power fraction is described by an equivalent-circuit view in which matching determines how much of the converted power reaches the load versus being dissipated internally (see Supplementary Information).

The outward-facing surface of the transducer is bonded to silver epoxy (MG 8330), acting as both a quarter-wave matching layer and an electrical connection. With an acoustic impedance of ~6 – 7 MRayl, silver epoxy is close to the ideal quarter-wave matching impedance $\sqrt{Z_{\text{tiss}} Z_{\text{tran}}} \sim 7.3$ MRayl needed to couple energy effectively between tissue ($Z_{\text{tiss}} = 1.5$ MRayl) and PZT-5A ($Z_{\text{tran}} = 36$ MRayl). The inner surface is air-backed to increase mechanical resonance and to prevent unwanted energy leakage into the capsule interior. At resonance, the receiver has an electrical impedance of $3.78$ k$\Omega \angle -47.8°$ as measured with an LCR meter (Hioki 3532-50).

The received 500 kHz electrical signal must then be rectified and stored for pulsed laser operation. We employ a two-stage Cockcroft-Walton voltage-multiplying circuit (Figure 3a), selected for its stable input impedance while charging the capacitor bank. The input impedance of the generator is measured as $186.1 \, \Omega \angle -12.2°$ during charging. An LC matching network is inserted between the receiver and multiplier circuit to maximize power transfer.

Acoustic power is delivered using a flat 1.5-inch diameter 500 kHz transducer (Olympus V389-SU) driven by an arbitrary waveform generator (Siglent SDG2042X) and power amplifier (E&I 350L). We also electrically match the transmitting transducer ($150.3 \, \Omega \angle -78.2°$) to the 50 $\Omega$ power amplifier using an LC matching network. The flat transmitting aperture produces a broad axial region of near-uniform intensity near the Rayleigh length $D_{\text{ray}} \sim D^2/4\lambda \sim 12$ cm, reducing sensitivity to capsule positioning. Given the uncontrolled motion and in the GI tract, this could improve the reliability of received power delivery compared with a sharply focused transmitter. Here, we maximize the safety-limited acoustic intensity that is converted into electrical energy at the capsule rather than the efficiency of the entire acoustic power transfer link.

We then used this system to wirelessly charge the capacitor bank used for laser firing, where a linear increase in voltage is found during the charging period (Figure 3b). A Zener diode is also used to limit the rectified DC voltage to 15 V, where additional acoustic power received after the capacitor bank is charged to 15 V is dissipated through the Zener diode. Using a calibrated hydrophone (Onda HGL-0085), we measured the peak acoustic pressure as $p_{\text{pk}} = 210$ kPa, corresponding to a mechanical index of ~0.3. This results in a peak acoustic intensity of $p_{\text{pk}}^2/2Z_a \sim 1.47$ W/cm$^2$, so a maximum duty cycle of ~49% can be used while remaining within the safety standards.

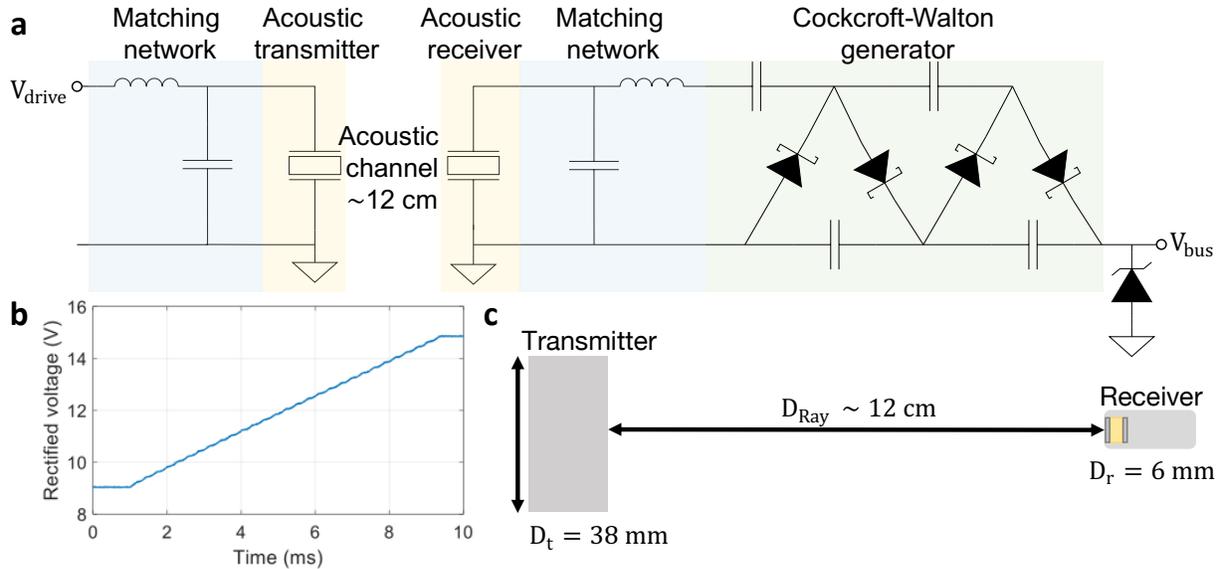

Figure 3. Acoustic power transfer for wireless photoacoustic tomography. **a** Simplified circuit diagram. $V_{drive}$ connects to a power amplifier driving the transmitter. $V_{bus}$ connects to the capacitor bank used to power the laser diode array. **b** Rectified voltage $V_{bus}$ from acoustic power transfer on the capacitor bank. **c** Schematic of the transmitter and receiver geometry.

*Wireless triggering*

To control laser firing and to synchronize with photoacoustic detection, we use low-latency wireless triggering. Alternatively, an independent trigger source could be included inside the device, which is synchronized to the external detection circuitry. However, this may lead to drift over time and would not provide flexibility in the repetition rate used during acquisition. For low-latency triggering, an energy source with high propagation speeds and low loss in tissues is desirable.

We transmit a broadband electromagnetic pulse using a fast-rise-time electrical pulse (Olympus 5073PR) connected to a water-immersed dipole antenna. The rise time is approximately 2 ns, corresponding to pulse frequency content up to several hundred MHz. In this frequency range, electromagnetic waves exhibit moderate attenuation in aqueous and biological media, enabling penetration that is sufficient for the dimensions relevant to the gastrointestinal tract. The transmitted signals are detected using a conformal helical antenna surrounding the device operating in broadside mode. While a resonant antenna could be more efficient, the varying dielectric properties throughout the GI tract may unexpectedly alter the antenna behavior due to dielectric loading.

Since the trigger input is high impedance, the received energy can be small while generating reasonable voltages. The antenna is connected to a full-wave rectifier constructed using ultrafast diodes (SMS7621). A 10 kΩ pull-down resistor is placed at the output terminal to shorten the rise times associated with the internal capacitance of the diodes. The rectifier output then connects to a monostable multivibrator circuit to generate pulses of consistent duration. The resistor and capacitor values are chosen to generate ~0.6 µs pulses for triggering the laser diodes through the gate driver. An example wirelessly-generated pulse is shown in Figure 4. This module also contains a low-dropout voltage regulator to obtain a 5 V source from the 15 V obtained from acoustic power transfer.

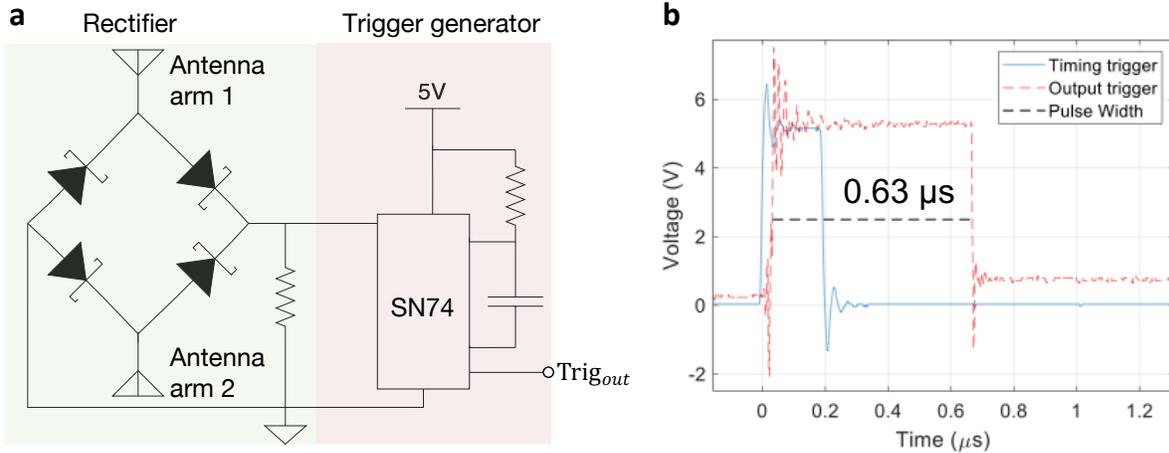

Figure 4. Wireless triggering circuit. **a** Simplified circuit diagram. SN74: monostable multivibrator integrated circuit. The pulse duration is fixed by the RC values used in the trigger generator circuit. **b** Recorded wirelessly triggered pulse.

**Results**

The device is housed in a 3D-printed capsule with a PDMS window on the front and rear surfaces. PDMS is used to provide a protective barrier that could survive the GI tract without substantially attenuating acoustic or optical energy. The receiving ultrasonic transducer is wired to the matching network, which then connects to the acoustic rectifier PCB. The two arms of the helical antenna are connected to the triggering PCB. Prior to encapsulation, each circuit is mounted using a 3D-printed holder. The three PCBs are connected mechanically and electrically using four lines: ground, 5 V, 15 V, and trigger.

The acoustically powered and electromagnetically triggered WPAT device is shown in Figure 5, and the device operation with varying position is shown in Supplementary Video 1. The device is mounted on an acoustic absorber to reduce reverberation in the water tank. These images were recorded using a scientific camera (FLIR GS3-U3-23S6M). Note that the camera's sensitivity at 905 nm is ~10 times weaker than at visible wavelengths.

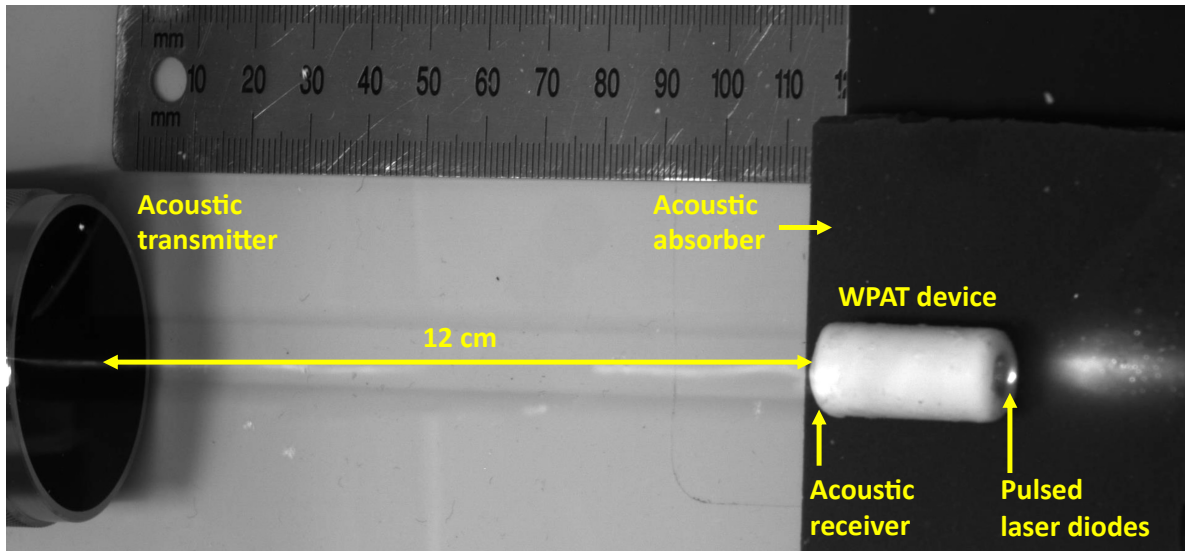

Figure 5. WPAT device operation with acoustic power transfer through 12 cm of water.

The system timing diagram for imaging is shown in Figure 6a. First, acoustic power transfer is performed over 8 milliseconds, charging the laser module capacitor banks. Due to the high acoustic intensity used for device powering and the low acoustic attenuation at 500 kHz, we use a 10 ms gap to allow the acoustic powering signals to decay so that they do not dominate the weak photoacoustic signals. Note that this time could likely be reduced in human tissue due to greater attenuation and scattering. The resulting photoacoustic signals are recorded using the custom 512-element, 60 cm diameter array described in [17]. Following detection, the next acoustic charging cycle begins. This results in an acoustic powering duty cycle of 40%, corresponding to an $I_{SPTA} \sim$ 590 mW/cm$^2$.

We demonstrate this approach by imaging a target while performing acoustic power transfer through a 12 cm thick 4% agar phantom, which we use as a test scenario for human GI imaging applications. We scan the transmitter and WPAT device laterally in two dimensions (Figure 6b) in 4 mm steps. At each position, we average the photoacoustic response from 2000 laser shots. We then apply a notch filter at 500 kHz on the recorded signals to reduce remaining reverberation from acoustic powering. The PAT images from each device position are reconstructed, and a 6 mm Gaussian window is used to isolate photoacoustic signals from the local optical excitation. We scan 45 positions to construct a cm-scale target image (Figure 6d). We emphasize that despite the very weak optical pulse energy (130 µJ) and human-scale distances from the acoustic transmitter and photoacoustic receivers we can generate PAT images using an entirely wireless device in a small form factor.

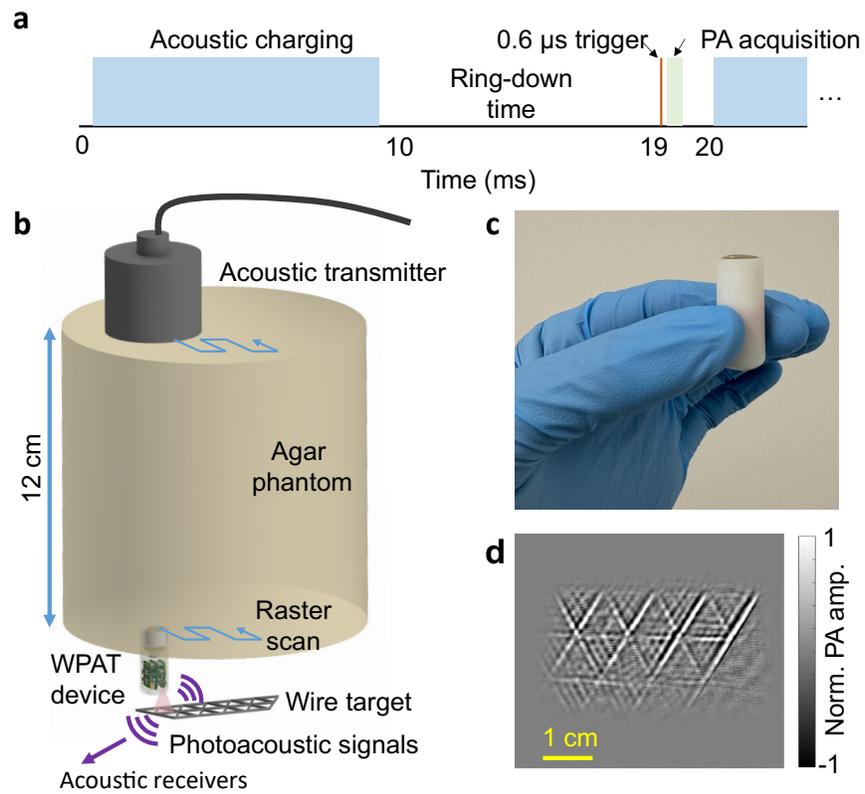

Figure 6. Experimental configuration and results for WPAT phantom imaging. **a** Timing diagram for acoustically powering the device and recording PA signals. **b** Experimental configuration for imaging through a 12 cm phantom. **c** Assembled device. **d** Example generated PAT image.

## Discussion

We have demonstrated wireless PAT at clinically relevant depths using a wirelessly powered and triggered device in an ingestible form factor. This technology may enable imaging through mucosal layers in the entire GI tract, which is not possible with any existing approach. Acoustic power transfer eliminates the need for onboard batteries, which could enable prolonged continuous imaging during natural transit through the GI tract. In the future, large animal studies will be important to evaluate WPAT's clinical utility, including the detection of inflammatory lesions in Crohn's disease or the characterization of submucosal tumors.

A remaining challenge for clinical translation is the potential presence of air pockets in parts of the GI tract, which could impede both acoustic power transfer and photoacoustic signal detection. While the small intestine is generally filled with fluid [27], the stomach and large intestine often contain larger volumes of air. Patient preparation protocols may therefore be necessary to reduce acoustic shadowing. Additionally, although we employed a single-element transmitter for wireless powering, heterogeneous abdominal tissue may require adaptive beamforming to efficiently deliver acoustic energy to the device. Our experiment also assumes a known device location, but in the realistic GI tract the device will change position and orientation during transit. Robust localization strategies using, for instance, magnetic fields [28], [29], electromagnetic waves [30], or ultrasound imaging could enable real-time device tracking for optimal acoustic beam steering to power the device.

Since image acquisition is performed over several seconds, motion artifacts *in vivo* will need to be accounted for. Conformal ultrasound arrays [31], which may become commercially available in the future, could be wrapped around the waist to enable continuous imaging and device powering throughout the GI tract. While these could compensate for bulk patient motion, differential movement between the skin and abdominal organs must also be considered. Several additional features of this device could be optimized for GI imaging. For example, in conventional CE, side-viewing optics from the device may provide more clinically useful images than forward-looking views [32]. Future WPAT designs could therefore incorporate lateral optical illumination to better image structures perpendicular to the axis of transit.

Other mechanisms, such as sonoluminescence [33], [34] or microwave-driven plasma, can also convert other forms of energy into optical emission, but they face key limitations for biomedical use. The acoustic pressures required to generate sonoluminescence far exceed human safety thresholds, and the resulting optical source would be accompanied by strong nonlinear acoustic responses that could mask the weak photoacoustic signals. Similarly, generating microwave-induced plasma would require high electric field amplitudes and specialized gas environments that are likely infeasible for human use.

Future WPAT implementations could incorporate an onboard acoustic receiver to detect photoacoustic signals closer to the source. Local detection would allow the use of higher frequency ultrasound, improving spatial resolution with reduced attenuation and aberration through tissue. However, integrating front-end amplification and digitization would substantially increase the device's power requirements compared with the external photoacoustic detection demonstrated here. As an alternative, passive readout methods such as radiofrequency [35] or magnetoelectric

[36] backscatter links may enable the transmission of encoded acoustic information without requiring high-power electronics in the capsule.

**Acknowledgements**

This project has been made possible in part by grant number 2024-337784 from the Chan Zuckerberg Initiative DAF, an advised fund of the Silicon Valley Community Foundation. L.W. has a financial interest in Microphotoacoustics, Inc., CalPACT, LLC, and Union Photoacoustic Technologies, Ltd., which, however, did not support this work.

Supplementary Information for:

# An Ingestible Light Source for Deep Photoacoustic Imaging

David C. Garrett and Lihong V. Wang

*Caltech Optical Imaging Laboratory, Andrew and Peggy Cherng Department of Medical Engineering, Department of Electrical Engineering, California Institute of Technology, Pasadena, CA 91125, USA*

\* Corresponding author: lvw@caltech.edu

**Sensitivity analysis**

The optical energy of ~ 130 µJ used here is far lower than that in typical PAT systems, but we illuminate a narrower region near the laser. Here, we theoretically estimate the SNR of this system. While the fluence in this system depends on the target distance from the laser surface, we approximate it here as $F \sim 130 \text{ µJ/cm}^2$. For our pulse width of $\Delta t = 0.6$ µs, photoacoustic stress confinement is valid for imaging absorbers with characteristic size $d > \Delta t c_s \sim 0.9$ mm, where $c_s \sim 1540$ m/s is the speed of sound in tissue. The resulting initial photoacoustic pressure is $p_0 = \Gamma \mu_a F$ [1]. For tissue, we consider $\mu_a \sim 0.1 \text{ cm}^{-1}$, and for the wire target we approximate $\mu_a \sim 2.0 \text{ cm}^{-1}$. The Grüneisen coefficient $\Gamma$ is ~ 0.2 (dimensionless). This results in an initial pressure $p_{0,\text{tissue}} \sim 2.6$ Pa and $p_{0,\text{wire}} \sim 52$ Pa.

We then compare this pressure with the noise-equivalent pressure (NEP) of our system [2]. We consider an NEP spectral density of 0.5 mPa Hz$^{-1/2}$ of each acoustic receiver element [3], corresponding to NEP$_0$ as 0.5 Pa over a 1 MHz bandwidth. The NEP of a single recorded receiver channel NEP$_{\text{sig}}$ can be estimated by averaging over the number of shots $N_{\text{avg}}$ and by scaling the NEP from the receiver surface to the target location $r$. In our human-scale system, $r \sim 30$ cm. We consider the acoustic wavelength at 1 MHz, and our experiments used $N_{\text{avg}} = 2000$.

$$\text{NEP}_{\text{sig}} = \text{NEP}_0 \frac{r/\lambda}{\sqrt{N_{\text{avg}}}} \sim 2.2 \text{ Pa} \tag{1}$$

This would result in a signal-to-noise ratio (SNR) of $p_{0,\text{wire}}/\text{NEP}_{\text{sig}} \sim 24$. In the recorded signals shown in Supplementary Figure 1, we observe an SNR of approximately 27. Note that these signals were notch filtered at 500 kHz to remove residual signals from acoustic power transfer, but some ringing remains.

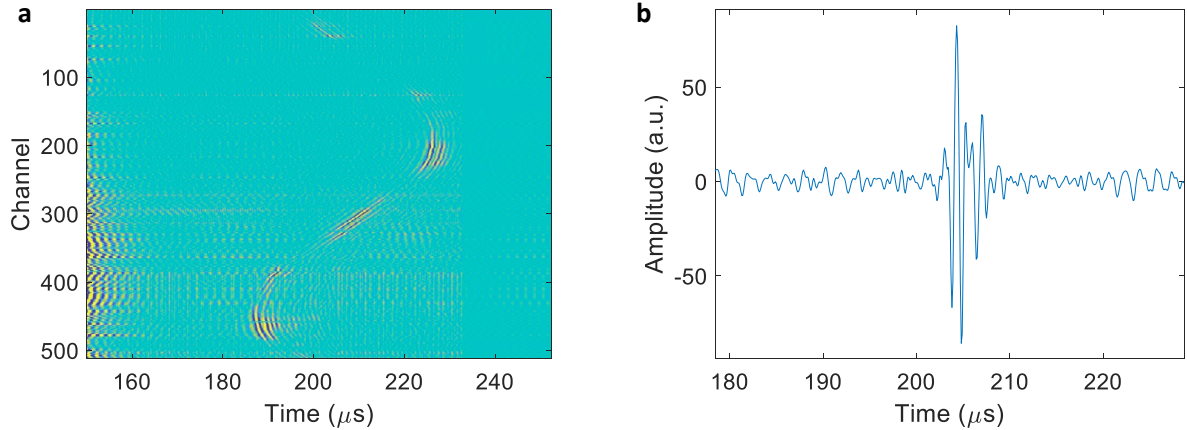

Supplementary Figure 1. Example recorded signals using WPAT excitation. **a** Recorded sinogram across all 512 channels after notch filtering the 500 kHz acoustic power transfer signal. **b** Example signal on channel 310.

The resulting NEP of the entire imaging system can be estimated [3] by also averaging over the receiver element count $N_{\text{ele}} = 512$. The resulting imaging $\text{NEP}_{\text{img}}$ is estimated as

$$\text{NEP}_{\text{img}} = \text{NEP}_0 \frac{r/\lambda}{\sqrt{N_{\text{avg}}}\sqrt{N_{\text{ele}}}} \sim 0.1 \text{ Pa} \qquad (2)$$

In realistic tissues with $\mu_a \sim 0.1 \text{ cm}^{-1}$, we expect the resulting image SNR to be approximately:

$$\text{SNR}_{\text{img}} = \frac{p_{0,\text{tissue}}}{\text{NEP}_{\text{img}}} \sim 26 \qquad (3)$$

## Image generation

We raster-scanned the WPAT device to generate a cm-scale image of a wire target. For each position, we reconstruct the photoacoustic image. We apply a Gaussian window with $\sigma = 6$ mm centered at the device position to isolate the illuminated imaging region while minimizing the inclusion of image noise. An example window and image for a single position is shown in Supplementary Figure *2*. The final image is the sum of the images from all 50 positions.

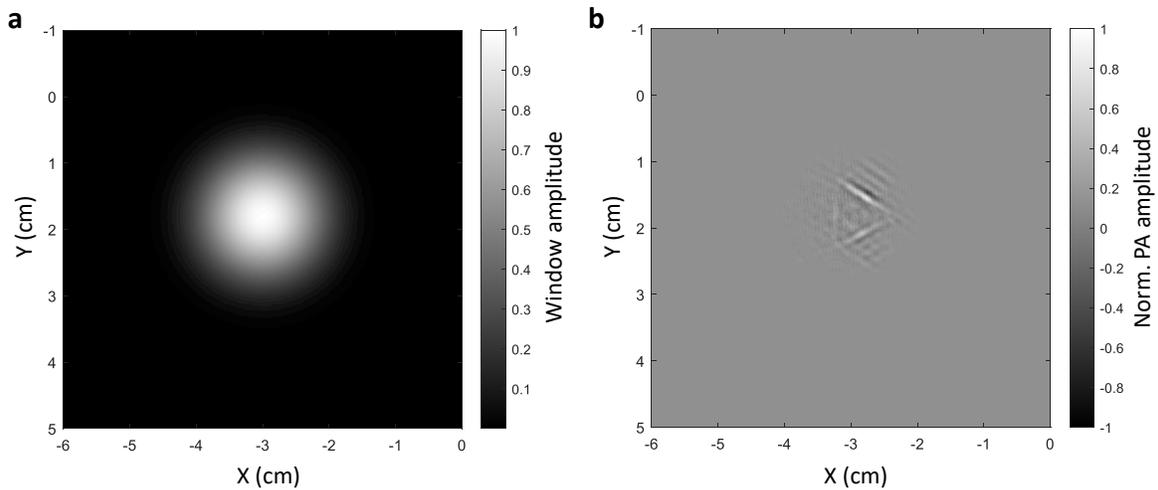

Supplementary Figure 2. Example image from a single device location during raster-scanning. **a** Gaussian window used to isolate the illuminated region. **b** Photoacoustic image after applying the Gaussian window. The final image is generated by summing the windowed images from all device positions.

## Acoustic power transfer

We use a flat 1.5-inch diameter transmitter to power the WPAT device. This geometry corresponds to a Rayleigh length of ~12 cm. Compared with focused transmitters, flat ones provide a more uniform focal region with reduced positioning precision requirements. The simulated beam profile of this transducer is shown in Supplementary Figure 3.

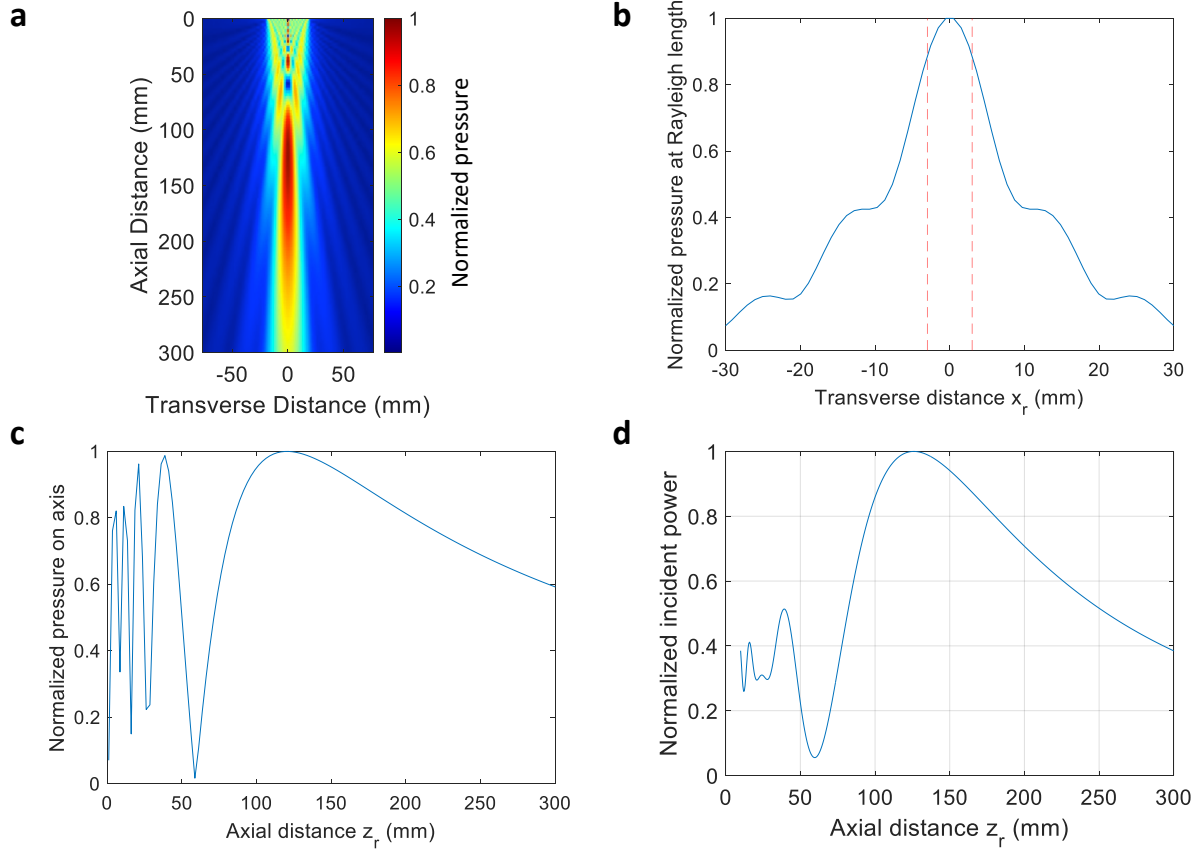

Supplementary Figure 3. Simulated acoustic beam profile for acoustic power transfer. **a** Simulated beam profile for the transmitter. **b** Transverse profile of the normalized pressure at the Rayleigh length of 12 cm. The red-dashed lines indicate the diameter of the receiving transducer. **c** Simulated on-axis normalized pressure. **d** Simulated on-axis power transferred to a receiving 6 mm diameter element.

Supplementary Figure 3d also shows the simulated on-axis normalized power transfer between the transmitting element and the 6 mm diameter receiver. Note that 80% of peak power transfer is maintained over a ~8 cm axial distance, which relaxes positioning accuracy requirements in our experiments.

We show the rectified DC voltage from acoustic power transfer to the capacitor bank in Supplementary Figure 4. Laser firing causes a sharp drop in voltage from ~15 V to 9 V, corresponding to an energy of $E_{LD} = \frac{1}{2}C(V_i^2 - V_f^2) \sim 324$ µJ. Note that this example is shown for 100 Hz laser repetition rate.

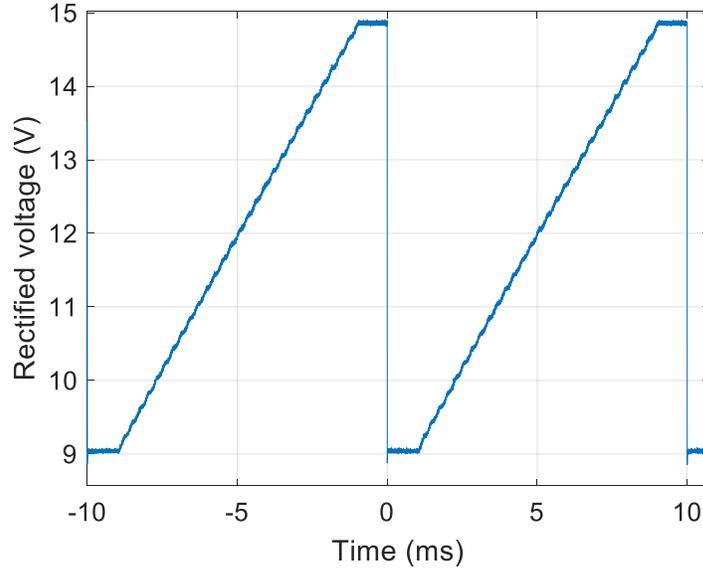

Supplementary Figure 4. Example rectified voltage from the acoustic receiver to the capacitor bank used to energize the pulsed laser diode array. The laser is fired here every 10 ms.

We estimate the efficiency of the receiver transducer based on the open-circuit recorded voltage and the incident pressure on the receiver surface. We used a hydrophone to record the incident peak pressure as $p_{\text{pk}} \sim 145$ kPa using a representative continuous-wave transmitter power. The available power at the receiver, assuming a uniform pressure distribution over the small receiver area $A$, is therefore

$$P_{\text{avail}} = \frac{(p_{\text{pk}})^2}{2Z_a} A \approx 198 \text{ mW}. \tag{4}$$

We then recorded the open-circuit voltage from the receiver as $V_{\text{oc}} \sim 30$ V, resulting in an estimated received power of

$$P_{\text{rx}} = \frac{(V_{\text{oc}})^2}{8\text{Re}\{Z_{\text{out}}\}} \approx 50 \text{ mW}, \tag{5}$$

where $\text{Re}\{Z_{\text{out}}\} \sim 2.54$ kΩ is the real part of the receiver output impedance. This results in an estimated transducer efficiency of

$$\eta = \frac{P_{rx}}{P_{avail}} \approx 26 \text{ \%} \tag{6}$$

We compare this with the KLM model [4] for PZT-5A, using $g_{33} \sim 0.025$ V · m/N and thickness $d = 4$ mm. The open-circuit voltage is estimated as:

$$V_{\text{oc}}(\omega) = g_{33} d T_{33,\text{eff}}(\omega) \tag{7}$$

where $T_{33,\text{eff}}$ is the effective normal stress in the piezoelectric, found from the incident acoustic pressure as $T_{33,\text{eff}}(\omega) = p_{\text{inc}} G_{\text{ac}}(\omega)$. The stress gain $G_{\text{ac}}(\omega)$ accounts for the quarter-wave matching layer and mechanical resonance from the air backing. Here, the measured $V_{\text{oc}} \sim 30$ V is achieved for $G_{\text{ac}}(\omega) \sim 2$, which is a modest value for an air-backed resonant transducer.

Note that after the matching circuit into the Cockcroft-Walton generator, the input voltage is lower due to the lower input impedance. Given the acoustic-to-electric efficiency from the transducer of ~26%, the required efficiency of the matching circuit and generator is ~31% to achieve a total required efficiency of 8% to meet our power requirements given the acoustic safety standards.

## Device construction

Steps of the device construction are shown in Supplementary Figure 5. The front surface of the 3D printed capsule is first coated with PDMS. The wired device, including the receiving transducer, matching network, and the three PCB modules, are then positioned into the capsule. The two arms of the antenna are connected to one of the PCBs. After insertion into the device, the rear face of the receiving transducer is also coated in a thin layer of PDMS.

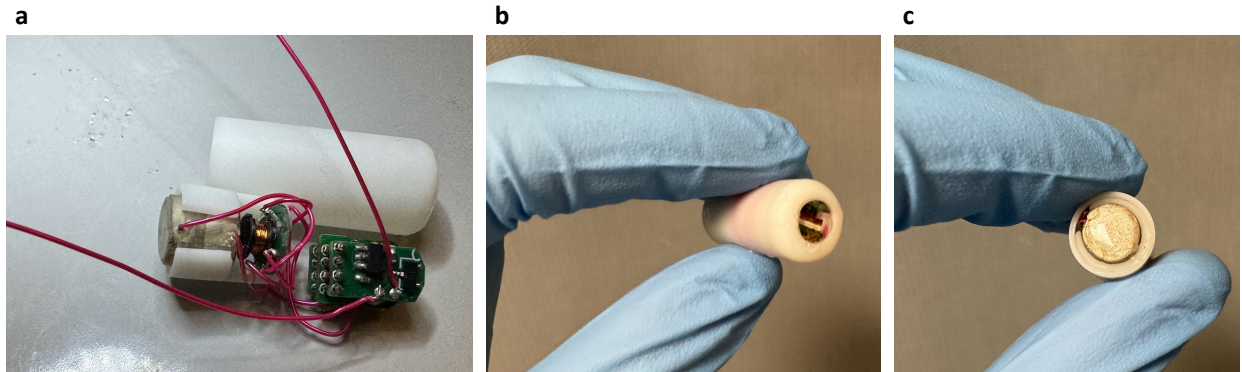

Supplementary Figure 5. Device construction. a Wired device before inserting into the 3D-printed enclosure. b Front face of the device with the laser diode array. c Rear face of the device with the receiving transducer. Both faces are coated with PDMS.